\documentstyle[twocolumn,aps]{revtex}

\begin{document}
\tightenlines
\title{The central density of  a neutron star is unaffected by \\ a binary
companion at linear order in $\mu/R$}
\author{Patrick R.\ Brady and Scott A.\ Hughes} 
\address{Theoretical Astrophysics, California Institute of Technology,
Pasadena, CA 91125}
\date{\today}
\maketitle
\draft
\begin{abstract}
Recent numerical work by Wilson, Mathews, and Marronetti [J.\ R.\
Wilson, G.\ J.\ Mathews and P.\ Marronetti, Phys.\ Rev.\ D {\bf 54},
1317 (1996)] on the coalescence of massive binary neutron stars shows
a striking instability as the stars come close together: Each star's
central density increases by an amount proportional to 1/(orbital
radius).  This overwhelms any stabilizing effects of tidal coupling
[which are proportional to 1/(orbital radius$)^6$] and causes the
stars to collapse before they merge.  Since the claimed increase of
density scales with the stars' mass, it should also show up in a
perturbation limit where a point particle of mass $\mu$ orbits a
neutron star.  We prove analytically that this does {\it not} happen;
the neutron star's central density is {\it unaffected} by the
companion's presence to linear order in $\mu/R$.  We show, further,
that the density increase observed by Wilson et.\ al.\ could arise as
a consequence of not faithfully maintaining boundary conditions.
\end{abstract}

Wilson, Mathews, and Marronetti (WMM)~\cite{WMM} have proposed a
method of approximating the fully General Relativistic analysis
of binary neutron star coalescence.  The essence of their
scheme is to choose a simple form of the spacetime metric (one in
which the spatial three slices are conformally flat), and solve the
constraint equations of General Relativity (GR) for some initial
matter configuration.  They evolve only the fluid equations forward in
time until the fluid reaches a quasi-equilibrium configuration, then
solve the constraint equations again for the new matter configuration
and iterate until a quasi-equilibrium solution to the combined
Einstein-fluid equations is found.  Their method makes 3-dimensional
simulations of such systems more tractable by reducing the
computational requirements.  

These simulations yield an extremely surprising result: neutron stars
that are close to the maximum allowed mass are ``crushed'' into black
holes long before the neutron stars coalesce.  They claim that
the origin of this effect is a non-linear gravitational interaction
due to the companion's presence that strengthens the gravitational
potential of each star.  Consider a binary star system---star-A has
mass $M_A$, and star-B has mass $M_B$.  WMM claim that non-linear
interactions cause the potential at star-A to be increased by a term
that scales as $M_B/R$ (where $R$ is the orbital separation).  This,
in turn, increases the internal energy and density of star-A by terms
that scale as $M_B/R$.  If star-A happens to be marginally stable in
isolation, the effect is sufficient to push it over the edge, causing
a catastrophic collapse to a black hole.

The scaling law claimed by WMM is precisely what one would expect if
this effect were due to a post-1-Newtonian enhancement of the
gravitational interaction.  Motivated by this observation, Wiseman has
recently done a careful analysis of the effect that a binary companion
has on a fluid star, using the first post-Newtonian approximation to
GR~\cite{alan}; he finds no change to either the central energy
density or the angle averaged proper radius of the star at this order.
Wiseman's calculation does not rule out completely a star-crushing
effect, but does show that it is not evident at post-1-Newtonian order
in GR.

Suppose for a moment that the WMM effect is a property of neutron star
binaries in GR, and that it scales as $M_B/R$ at star-A.  Clearly, it
should also be apparent in the limit that we shrink star-B down to a
point particle of mass $\mu \ll M_A\equiv M$.  In this limit, the
exact solution of the Einstein field equations describing a binary
neutron star system can be approximated by a perturbative expansion in
$\mu/R$ about the solution for an isolated star.  We write the metric
as $g_{\alpha\beta} = g^0_{\alpha\beta} + \epsilon h_{\alpha\beta} +
O(\epsilon^2)$, where the superscript $0$ indicates the background
metric, and we have introduced an order counting parameter $\epsilon$
with the formal value unity.  Quantities multiplied by $\epsilon$
scale linearly with $\mu/R$, quantities multiplied by $\epsilon^2$
scale with $(\mu/R)^2$, {\it etc.}  In what follows, we ruthlessly
discard all terms of order $\epsilon^2$, constructing an argument that
is valid only to linear order in $\mu/R$.

The neutron star material is considered to be perfect fluid with
stress-tensor
\begin{equation}
T_{\alpha\beta} = P\, g_{\alpha\beta} + (P +
	\rho) u_\alpha u_\beta \; .
\label{OVstress}
\end{equation}
This must be supplemented with an equation of state relating the
pressure $P$ and the energy density $\rho$.  For concreteness, we
assume a polytropic form $P= K n^\Gamma$ where $K$ and $\Gamma$ are
constants, and $n$ is the fluid's baryon density.  The energy density
$\rho$ is directly related to $n$ by the first law of thermodynamics;
see Eqs.~(3.2.6-7) of Ref.~\cite{shapandteuk}.

We take the background spacetime to be that of an isolated, spherical
star with the line element
\begin{equation}
ds^2 = - e^{2\Phi(r)} dt^2 + \frac{dr^2}{\left[1 - 2m(r)/r\right]} 
	+ r^2 d\Omega^2.
\label{sphline}
\end{equation}
Here $m(r)$ is the gravitational mass inside a sphere of radius $r$,
and $d\Omega^2 = d\theta^2 + \sin^2\theta\,d\phi^2$.  The combined
Einstein-perfect fluid equations, generally referred to as the
Oppenheimer-Volkoff (OV) equations (see for example Chapter 23 of
Ref.~\cite{MTW}) are solved by demanding regularity of the origin
[$m(0)=0$] and fixing the value of the central baryon density $n_c$.
The radius of the star $R_S^{\mbox{\scriptsize 0}}$ is the
coordinate radius at which the baryon density $n^{\mbox{\scriptsize
0}}$ becomes zero.  The central density $n_c$ uniquely determines
$R_S^{\mbox{\scriptsize 0}}$, the total mass
$M=m(R_S^{\mbox{\scriptsize 0}})$, and baryon mass $M_b$; this
statement is equivalent to Theorem~7 of Ref.~\cite{HTWW}.

One can show using the polytropic equation of state and the OV
equation for the pressure,
\begin{equation}
{dP^0\over dr} = -{(\rho^0 + P^0)[m(r) + 4\pi r^3 P^0]\over r[r-2m(r)]} \; ,
\end{equation}
that
\begin{equation}
\left.{dn^{0}\over dr}\right|_{r\to0} \to 0 \; . \label{dndrcondition_1}
\end{equation}
Finally, the background geometry outside the star is described by the
Schwarzschild solution with $m=M$ and $\exp(2\Phi)= 1 - 2M /r$ in
Eq.~(\ref{sphline}).
 
The perturbing source is a single point particle of proper mass $\mu$
in a circular orbit at radius $R$.  It is described by the
stress-energy tensor~\cite{eric}
\begin{equation}
T^{\alpha\beta} = {\epsilon\mu\over R^2}{v^\alpha v^\beta\over
	v^t}\delta(r-R)\delta(\cos\theta)\delta(\phi-\Omega t),
\label{particlestress}
\end{equation}
where $v^\alpha = (1 - 3M/R)^{-1/2} (1,0,0,\Omega)$ and $\Omega =
\sqrt{M/R^3}$.  The presence of this point ``star'' will alter the
geometry and disturb the material in the central star, modifying the
description of the spacetime and matter by terms of order $\epsilon$.
Linearizing $G_{\alpha\beta} = 8\pi T_{\alpha\beta}$ and
${T^{\alpha\beta}}_{;\beta}=0$ in $\epsilon$, we find that the
first-order perturbation equations separate by expanding the angular
dependence in spherical harmonics, and the time dependence in Fourier
modes.  This is enough to address the issue of how the central density
scales.

Consider the expansion of the baryon mass density.  It may be written
\begin{equation}
n(r,\theta,\phi,t) = n^{\mbox{\scriptsize 0}}(r) +
\epsilon\!\!\sum_{l,m,\omega}
\delta n_{lm\omega}(r) Y_{lm}(\theta,\phi) e^{i\omega t} \; .
\label{rhoexpand}
\end{equation}
An immediate consequence of Eq.~(\ref{rhoexpand}) is that $\delta
n_{lm\omega}(0) = 0$ for $l \ge 1$: if it were non-zero, the density
would be multi-valued at $r=0$.  Thus, only the monopole could affect
the central density if the center of the perturbed star were to remain
at the origin.  In reality, the star's center star will not be at the coordinate
origin: the orbiting body will move it to some point in the orbital plane.
However, the magnitude of this shift must be of the same order as the
perturbation itself: $r_{\rm cent} = \epsilon{\bf\xi}(t)$ for some function
$\xi(t)$.  Now evaluate the density at the star's center with a Taylor
expansion:
\begin{eqnarray}
&n_{\rm cent}& = n^{\mbox{\scriptsize 0}}(0) + \epsilon\xi 
	\left.{dn^{\mbox{\scriptsize 0}}\over dr}\right|_{r=0} \nonumber\\
        &+&\!\! \epsilon\sum_{l,m,\omega}\left[\delta n_{lm\omega}(0) +
	\epsilon\xi\left.{d\delta n_{lm\omega}\over dr}\right|_{r=0}
	\right] Y_{lm}(\theta,\phi)e^{i\omega t} \; .
\label{rhocomI}
\end{eqnarray}
As we have already shown, $\delta n_{lm\omega}(0)=0$, except possibly
for $l=0$, while $d n^{\mbox{\scriptsize 0}}/dr\to 0$ as $r\to0$ by
Eq.~(\ref{dndrcondition_1}).  Thus, the baryon density at the center
of mass is given by
\begin{equation}
n_{\rm cent} = n^{\mbox{\scriptsize 0}}(0) + \epsilon\delta n_{000}(0) +
	O(\epsilon^2)\; .
\label{rhocomII}
\end{equation}
{\em Only the monopole can produce changes in the central density
which scale linearly in $\mu/R$.}

It is straightforward to solve for the $l=m=0$ corrections to the
metricoutside the fluid.  Define the function
\begin{equation}
	H(r) =
\left\{ \begin{array}{lr}
	0 & r<R \\
	\displaystyle \frac{2 \mu}{r}\frac{(1-2M/R)}{(1-3M/R)^{1/2}} & \ \ r>R
	\end{array} \right. \; ;
\end{equation}
then $h_{tt} = H(r)$, $h_{rr} = H(r)/(1 - 2M/r)^2$.  We have set $h_{\theta\theta}
= h_{\phi\phi} = 0$ using the first
order gauge freedom available for the monopole.  As the point particle
spirals into the star, some of its total mass-energy (its contribution
to the total mass as measured at infinity) is radiated away, though
its locally measured mass (rest mass) is conserved.  The
multiplicative factor in the above expression correctly accounts for
that radiation loss.  Notice that there is no monopole contribution to
the metric inside the orbital radius of the particle---Keplerian
orbits inside the orbital radius measure {\it only} the mass of the
unperturbed neutron star at monopole order.

Is it possible for the monopole part of the perturbation to rearrange
the fluid in the star, but leave its total gravitational mass
unchanged?  The answer is unequivocally no.  A monopole perturbation
is spherically symmetric and can only take one spherical solution into
another.  However, when the equation of state is fixed, all spherical
solutions are parameterized by the gravitational mass---for each value
of the gravitational mass $M$ there exists a unique spherical
configuration of the star.  Spherical solutions therefore exhibit a
one-to-one correspondence between the gravitational mass and the
central density of the star.  Since the gravitational mass $M$ is
unchanged at monopole order, the central density cannot be affected
either.  Hence, the central energy density of a neutron star is
unaffected by a binary companion at order $\mu/R$.  {\it There is no
crushing effect which scales linearly with $\mu/R$.}

Interestingly, it is easily shown that incorrectly imposing
boundary conditions can lead to an increase in central density at
order $\mu/R$.  If the total gravitational mass of the star {\it and
particle} is held fixed in a sequence of quasi-equilibrium solutions
(ignoring the gravitational radiation that causes the orbital radius
to shrink), {\it and} the particle's locally measured mass (rest
mass) is held fixed, then obviously the star's total mass and baryon
mass must go up by an amount of order $\mu M/R$, contrary to how a
real binary would behave.  This mass increase will drive the central
density up by a fractional amount of order $\mu/R$, which is what the
WMM simulations show.  We have no evidence that this is what actually
happens in the WMM simulations; it merely illustrates one way in which
the observed density increase could arise.  

It is a pleasure to thank Alan Wiseman for valuable conversations and
many helpful insights.  We are also grateful to Kip Thorne for
suggesting improvements in the presentation.  This work was supported
in part by NSF grants AST-9417371 and PHY--9434337.  SAH gratefully
acknowledges the support of a National Science Foundation Graduate
Fellowship.  PRB is supported by a PMA Division Prize Fellowship at
Caltech.


\begin{references}

\bibitem{WMM} J.\ R.\ Wilson and G.\ J.\ Mathews, Phys.\ Rev.\ Lett.\
	{\bf 75}, 4161 (1995); J.\ R.\ Wilson, G.\ J.\ Mathews and P.\
	Marronetti, Phys.\ Rev.\ D {\bf 54}, 1317 (1996); G.\ J.\ Mathews and
	J.\ R.\ Wilson, astro-ph/9701142.

\bibitem{alan} A.\ G.\ Wiseman, Phys.\ Rev.\ Lett., submitted.

\bibitem{MTW} C.\ W.\ Misner, K.\ S.\ Thorne, and J.\ A.\ Wheeler,
	{\it Gravitation} (Freeman, San Francisco, 1973).

\bibitem{shapandteuk} S.\ L.\ Shapiro and S.\ A.\ Teukolsky, {\it
	Black Holes, White Dwarfs, and Neutron Stars: The Physics of
	Compact Objects} (Wiley, New York, 1983).

\bibitem{HTWW} B.\ K.\ Harrison, K.\ S.\ Thorne, M.\ Wakano, and
	J.\ A.\ Wheeler, {\it Gravitation Theory and Gravitational
	Collapse} (University of Chicago Press, Chicago, 1965).

\bibitem{eric} E.\ Poisson, Phys.\ Rev.\ D {\bf 47}, 1497 (1993).

\end{references}
\end{document}